\documentclass[twocolumn,nofootinbib]{revtex4}%

\usepackage{amssymb}
\usepackage{amsfonts}
\usepackage{amsmath}
\usepackage[latin1]{inputenc}
\usepackage{graphicx,color}
\usepackage{slashed}
\usepackage{young}
\usepackage[vcentermath]{youngtab}

\makeatletter

\newcommand{\be}{\begin{equation}}
\newcommand{\ee}{\end{equation}}
\newcommand{\bea}{\begin{eqnarray}}
\newcommand{\eea}{\end{eqnarray}}

\begin{document}

\title{A deformation of the Curtright action}
\author{Sergio H\"ortner}
\affiliation{Centro de Estudios Cient\'{\i}ficos (CECs), Valdivia, Chile}

\begin{abstract}
We present a deformation of the action principle for a free tensor field of mixed symmetry (2,1) --the Curtright action, a dual formulation of five-dimensional linearized gravity. It is constructed as the dual theory of the Einstein-Hilbert action linearized around a de Sitter background, and its derivation relies on the use of the two-potential formalism as an intermediate step. The resulting action principle is spatially non-local and space-time covariance is not manifest, thus overcoming previous no-go results.

\end{abstract}

\maketitle


\section{Introduction}
\setcounter{equation}{0}














The construction of interacting gauge theories involving fields of mixed symmetry remains largely an open subject of research: although free theories can be constructed solely on the basis of the gauge principle \cite{Curtright}-\cite{Aulakh}, much is yet to be understood concerning interactions. There are two main areas of research where this type of tensor fields appear, and therefore the problem of deformations is relevant. One is the construction of interacting higher spin gauge theories, in relation with the mixed symmetry representations found in the infinite tower of higher spin massive excitations of the string theory spectrum. The other, which constitutes the subject of the present work, is the study of gravitational (or, more generally, higher spin) duality and hidden symmetries of gravitational theories.
\medskip

There is a significant amount of indications \cite{West2001}-\cite{Lambert} in support of the conjecture \cite{Julia1980}-\cite{Julia1997} that some extended Kac-Moody algebra underlies, as a larger symmetry of the unreduced theory, the well-established hidden symmetries of gravitational theories that appear upon toroidal compactification \cite{Ehlers0}-\cite{Cremmer2}. Within this picture, all the bosonic fields and their Hodge duals enter algebraically on an equal footing and are expected to be related among themselves by symmetry transformations of a highly non-trivial form. This would hold in particular for the graviton, with the peculiarity of being the only field in the bosonic sector of eleven dimensional supergravity whose Hodge dual, represented by a two-column $(D-3,1)$ Young diagram, is described by a mixed symmetry tensor. Nevertheless, it is precisely this mixed symmetry character of the dual graviton the origin of the difficulties encountered in completing the proof of the conjecture. A particularly delicate impasse one meets in this regard is found in \cite{HenBou}: a no-go result that forbids, under the hypotheses of locality and manifest space-time covariance, deformations of the free action for a tensor field of (2,1) Young type, originally proposed by Curtright as a dual description of linearized gravity in five dimensions \cite{Curtright}. These difficulties may be conceptually interpreted as a consequence of the lack of a notion of diffeomorphism covariance for mixed symmetry tensors. 

\medskip

The question we address here is whether it is possible to construct, by making use of the prepotential formalism \cite{DeserTeitel}, a non-local theory, the dual of five-dimensional Einstein-Hilbert action linearized around some curved background metric, that one could regard as a deformation of the Curtright action --namely, in the flat limit one should recover the Curtright action and its gauge symmetries. We find that the answer is affirmative, at least, when the linearization is performed around a de Sitter background. The reason that justifies our focus on de Sitter space-time,  as well as the election of the prepotential formalism, deserves some words of explanation. One first notices that de Sitter space-time admits a spatially flat slicing in an appropriate coordinate frame. This observation is crucial in order to extend the $SO(2)$ electric-magnetic duality invariance of the four-dimensional Pauli-Fierz theory \cite{HT} to the case where a positive cosmological constant is present \cite{Julialambda}. In higher dimensions, on the other hand, it has been studied how to relate the Pauli-Fierz action to the dual theory using the prepotential formulation as an intermediate step \cite{me5}. Thus, it seems opportune to investigate the prepotential formulation of the five-dimensional Einstein-Hilbert action linearized around de Sitter space-time --expressed in flat slicing coordinates-- with the purpose of constructing the corresponding deformation of the Curtright theory. We find that, working in a convenient gauge in which the linearized constraints take the same form as they do in the absence of a cosmological constant, one is able to construct such a deformation from the prepotential formulation of the action principle.  

\medskip

The rest of the article is organized as follows. In Section II we first describe the linearization of the Einstein-Hilbert action around a de Sitter background in any number of dimensions of the space-time manifold, as well as the canonical form of the theory; then we focus on five dimensions and solve the constraints in terms of two prepotentials, which turn out to retain the mixed symmetries found when the linearization is performed around a Minkowski background. After realizing in Section III that upon a particular gauge choice the constraints and the action take essentially the same form as in a Minkowski background, we construct the dual, non-local theory by introducing a pair of mixed symmetry canonical variables using the inversion formulae already known from the flat case. In Section IV we outline our conclusions and discuss the relevance of our result.





\section{Prepotetial formulation}

\subsection{Linearized action}

In the presence of a cosmological constant, the ADM form \cite{ADM}-\cite{Abbott} of the Einstein-Hilbert action principle for a metric $g_{\mu\nu}$ defined on a space-time manifold of arbitrary dimension $D$ reads

\be
S=\int d^{D}x\;\left[\pi^{ij}\dot{g}_{ij}-N{\cal{H}}-N_{i}{\cal{H}}^{i}\right],\label{actionADM2}
\ee
with
\bea
&&{\cal{H}}=-g^{1/2}(R-2\Lambda)-g^{-1}(\frac{\pi^2}{D-2}-g_{ik}g_{jl}\pi^{ij}\pi^{kl})\nonumber\\
\eea
and
\bea
&&{\cal{H}}^{i}=-2\pi^{ij}_{\ \ |j}.
\eea
The functions $N=(-g^{00})^{-1/2}$ and $N_{i}=g_{0i}$ denote respectively the lapse and the shift, whereas $g$ stands for the determinant of the induced metric in $D-1$ dimensions, $g=det(g_{ij})$. 


\medskip

In an arbitrary dimension $D$, the de Sitter line element expressed in spatially flat slicing coordinates takes the form
\be
ds^{2}=-dt^{2}+f^{2}(t)\delta_{ij}dx^{i}dx^{j}\label{metricdesitter}
\ee
with the definitions $f(t)=e^{kt}$ and $k=\sqrt{\frac{2\Lambda}{(D-1)(D-2)}}$. The linearization around this de Sitter background is achieved by setting
\bea
g_{ij}=\bar{g}_{ij}+h_{ij}, \ \ \ \pi^{ij}=\bar{\pi}^{ij}+p^{ij} \nonumber\label{expansionh} \\
N=1+n, \ \ \ N_{i}=n_{i},
\eea
where the barred quantities are determined from (\ref{metricdesitter}):
\bea
\bar{g}_{ij}&=&f^{2}(t)\delta_{ij}\nonumber\\
\bar{\pi}^{ij}&=&\sqrt{\bar{g}}(\bar{g}^{ij}\bar{K}-\bar{K}^{ij})=-(D-2)kf^{D-3}\delta^{ij}.\nonumber
\eea
By expanding (\ref{actionADM2}) up to second order terms in the graviton variables one finds the linearized version of the action,
\be
S=\int d^{D}x\left[p^{ij}\dot{h}_{ij}-H-nC-n_{i}C^{i}\right]\label{actionlambda2}
\ee
with the Hamiltonian density
\bea
&&H=f^{-D+5}p_{ij}p^{ij}-\frac{f^{-D+5}}{D-2}p^{2}-2(D-3)kp_{ij}h^{ij}+khp \nonumber\\
&&+f^{D-7}\left[\frac{1}{4}\partial^{i}h^{jk}\partial_{i}h_{jk}-\frac{1}{4}\partial_{i}h\partial^{i}h+\frac{1}{2}\partial^{i}h\partial^{j}h_{ij}\right.\nonumber\\
&&\left.-\frac{1}{2}\partial_{i}h^{ij}\partial^{k}h_{kj}\right]-k^{2}f^{D-5}\frac{(D-2)(-2D+6)}{4}h_{ij}h^{ij}\label{hamdensty}\nonumber\\
\eea
and the linearized constraints
\bea
C&\equiv&f^{D-5}(\Delta h-\partial_{i}\partial_{j}h^{ij})+2kpf^{2}\nonumber\\
&&+f^{D-3}k^{2}h(D-2)(D-3)=0\\
C^{i}&\equiv&-2\partial_{j}p^{ij}\nonumber\\
&&+(D-2)f^{D-5}k(2\partial_{k}h^{ik}-\partial^{i}h)=0
\eea
The latter are first-class and, as such, generate the gauge transformations of the canonical variables via the Poisson brackets
\bea
\delta h_{ij}&=&\{h_{ij}, \int d^{D-1}x (-\xi C+\xi_{m}C^{m})\}\nonumber\\
&=&\partial_{i}\xi_{j}+\partial_{j}\xi_{i}-2kf^{2}\delta_{ij}\xi\label{Poiss}\nonumber\\
\delta p_{ij}&=&\{p_{ij}, \int d^{D-1}x(-\xi C+\xi_{m}C^{m})\}\nonumber\\
&=&f^{D-5}(-\partial^{i}\partial^{j}\xi+\delta_{ij}\Delta\xi)\nonumber\\
&&+(D-2)(D-3)k^{2}f^{2}\delta_{ij}\xi\nonumber\\
&&+(D-2)k(\partial_{i}\xi_{j}+\partial_{j}\xi_{i}-\partial_{m}\xi^{m}\delta_{ij}).\nonumber\\
\eea
One should notice that the Hamiltonian and the kinetic term in (\ref{actionlambda2}) are not gauge invariant (up to a total derivative) by themselves, contrarily to the case $\Lambda=0$. Instead, the variation of the former compensates the variation of the latter. The reason for this is the explicit dependence of the Hamiltonian and the constraints on the time-like variable $t$. This forces the introduction of explicit time derivatives in the equations expressing the conservation in time of the constraints \cite{HenTei}, which otherwise would only involve brackets:\footnote{As described in \cite{HenTei}, in order to eliminate explicit time derivatives one may promote the time-like coordinate $t$ to a canonical variable by a reparametrization $t=t(\tau)$. The  phase space is then enlarged to contain both $t$ and its canonical momentum.}

\bea
\dot{C}&=&\frac{\partial C}{\partial t}+\{C,\int d^{D-1}y\;H\}=0\\
\dot{C^{m}}&=&\frac{\partial C ^{m}}{\partial t}+\{C^{m},\int d^{D-1}y\;H\}=0.
\eea







\subsection{Solving the constraints}

In order to solve the constraints, it is most useful to perform the canonical transformation

\bea
h_{ij}&\mapsto&\hat{h}_{ij}=h_{ij}\nonumber\\\label{cantra}
p^{ij}&\mapsto&\hat{p}^{ij}=p^{ij}-\frac{D-2}{2}kf^{D-5}(2h^{ij}-\delta^{ij}h).\nonumber\\
\eea
This transformation may be derived from the generating functional

\bea
F[h_{ij},\hat{p}^{ij}]&=&\int d^{D-1}x\left[\hat{p}^{ij}h_{ij}\right.\nonumber\\
&&\left.+\frac{(D-2)}{2}kf^{D-5}(h_{ij}h^{ij}-\frac{1}{2}h^{2})\right]\nonumber\\\label{genfunc}
\eea
depending on the `old' field $h_{ij}$ and the `new' conjugate momentum $\hat{p}^{ij}$, for it reproduces\footnote{For a discussion on generating functions in classical mechanics and how they are used to define the `old' and `new' sets of canonical variables, see for instance \cite{Landau}.} (\ref{cantra}):

\bea
\frac{\delta F}{\delta h^{ij}}&\equiv& p_{ij}=\hat{p}_{ij}+\frac{D-2}{2}kf^{D-5}(2h_{ij}-\delta_{ij}h)\nonumber\\
\frac{\delta F}{\delta \hat{p}^{ij}}&\equiv& \hat{h}_{ij}=h_{ij}.\nonumber
\eea
The action principle (\ref{actionlambda2}) reduces then to
\be
S[\hat{p}^{ij},h_{ij},n,n_{i}]=\int d^{D}x\left[\hat{p}^{ij}\dot{h}_{ij}-H-nC-n_{i}C^{i}\right],\label{actionlambda1}
\ee
where now the Hamiltonian and the constraints read, respectively,
\bea
&&H=f^{-D+5}\hat{p}_{ij}\hat{p}^{ij}-\frac{f^{-D+5}}{D-2}\hat{p}^{2}+2k\hat{p}_{ij}h^{ij} \nonumber\\
&&+f^{D-7}\left[\frac{1}{4}\partial^{i}h^{jk}\partial_{i}h_{jk}-\frac{1}{4}\partial_{i}h\partial^{i}h+\frac{1}{2}\partial^{i}h\partial^{j}h_{ij}\right.\nonumber\\
&&\left.-\frac{1}{2}\partial_{i}h^{ij}\partial^{k}h_{kj}\right]\label{Hamii}
\eea
and
\be
C=f^{D-5}(-\partial^{i}\partial^{j}h_{ij}+\Delta h)+2kf^{2}\hat{p}\label{Cscalar}
\ee
\be
C^{i}=-2\partial_{j}\hat{p}^{ij}.\label{vc}
\ee
We notice a dependence on the dimension of space-time $D$ in the scalar constraint, whereas the vector constraint does not depend on $D$ at all. The new canonical variable $\hat{p}^{ij}$ transforms as
\be
\delta \hat{p}^{ij}=f^{D-5}(-\partial^{i}\partial^{j}\xi+\delta^{ij}\Delta\xi).
\ee
In the sequel we shall focus on a five-dimensional space-time, where the dual graviton is described by the Curtright field, i.e., a mixed symmetry tensor field of Young type $(2,1)$. We shall solve the constraints in terms of prepotentials and write down the corresponding action principle. 

\medskip

The solution of the vector constraint (\ref{vc}) has been found in \cite{me5} by making use of the cohomological results for $N$-complexes and the generalized Poincar\'e lemma for rectangular Young diagrams \cite{DVH}: 

\be
\hat{p}^{ij}=\epsilon^{iklm}\epsilon^{jnpq}\partial_{k}\partial_{n} P_{lmpq}.\label{aeps}
\ee
The prepotential $P_{abcd}$ is a mixed symmetry tensor of Young type $(2,2)$. The transformations
\bea
\delta P_{abcd}&=&2\chi_{cd[b,a]}+2\chi_{ab[d,c]}+\frac{1}{4}[\delta_{ac}\delta_{bd}-\delta_{ad}\delta_{bc}]\xi\label{invPP},\nonumber\\ \label{gaugeP}
&&\chi_{abc}=-\chi_{bac}, \chi_{[abc]}=0
\eea
reflect the ambiguity in the choice of $P_{abcd}$: they leave $\hat{p}^{ij}$ invariant up to the gauge transformation $\delta \hat{p}^{ij}=-\partial^{i}\partial^{j}\xi+\delta^{ij}\Delta\xi$. Thus, the parameter $\xi$ in (\ref{gaugeP}) produces the gauge transformations of $\hat{p}^{ij}$, whereas $\chi_{abc}$ defines an internal invariance. 

\medskip

On the other  hand, the expansion of the product of the totally antisymmetric tensors in (\ref{aeps}) yields

\bea
\hat{p}^{ij}&=&\delta^{ij}(2\Delta P^{ab}_{\ \ ab}-4\partial_{a}\partial_{b}P^{amb}_{\ \ \ \ m})-2\partial^{i}\partial^{j}P^{ab}_{\ \ ab}\nonumber\\
&&+4\partial^{i}\partial_{a}P^{jba}_{\ \ \ \ b}+4\partial^{j}\partial_{a}P^{iba}_{\ \ \ \ b}\nonumber\\
&&-4\Delta P^{iaj}_{\ \ \ \ a}+4\partial_{a}\partial_{b}P^{iajb}
\eea
so one gets from its trace and (\ref{cantra})
\be
p=-3kh+2\Delta P^{ab}_{\ \ ab}-4\partial_{a}\partial_{b}P^{amb}_{\ \ \ m}.
\ee
Substitution in the scalar constraint (\ref{Cscalar}) produces

\be
\Delta h-\partial_{i}\partial_{j}h^{ij}+4f^{2}k\Delta P^{ab}_{\ \ ab}-8f^{2}k\partial_{i}\partial_{j}P^{imj}_{\ \ \ \ m}=0.\label{Clam22}
\ee
In order to solve the previous equation we shall decompose the prepotential $P_{ijkl}$ as follows:

\be
P_{abcd}=Q_{abcd}+\frac{1}{12}\left[\delta_{ac}\delta_{bd}-\delta_{ad}\delta_{bc}\right]P_{mn}^{\ \ mn}\label{PQ}
\ee
with $Q_{ijkl}$ a (2,2) tensor whose double trace vanishes. According to (\ref{gaugeP}), the term carrying the double trace of (\ref{PQ}) is a gauge transformation of parameter $\xi=P^{mn}_{\ \ mn}/3$. This transformation also affects $h_{ij}$ through (\ref{Poiss}) but not the form of the scalar constraint, owing to its gauge invariance. Equation (\ref{Clam22}) may then be written in the form

\be
\Delta j-\partial^{i}\partial^{j}j_{ij}+4kf^{2}\Delta Q^{ab}_{\ \ ab}-8kf^{2}\partial^{i}\partial^{j}Q_{ikj}^{\ \ \ k}=0
\ee 
with $h_{ij}=j_{ij}-\frac{2}{3}kf^{2}\delta_{ij}P^{mn}_{\ \ mn}+\partial_{i}u_{j}+\partial_{j}u_{i}$. One may choose $u^{i}$ in such a way that $j=0$ so the scalar constraint reduces to

\be
\partial^{i}\partial^{j}(j_{ij}+8kf^{2}Q_{ikj}^{\ \ \ k})=0.
\ee
This equation can be solved exactly in the same manner as the Hamiltonian constraint in the case $\Lambda=0$ \cite{me5}:

\be
j_{ij}+8kf^{2}Q_{ikj}^{\ \ \ k}=\partial^{k}\epsilon_{ikab}\phi^{ab}_{\ \ j}+\partial^{k}\epsilon_{jkab}\phi^{ab}_{\ \ i}.\label{leftside}
\ee
The traceless condition on the left-hand side of (\ref{leftside}) guarantees that the prepotential $\phi_{ijk}$ is a mixed symmetry tensor transforming in the (2,1) irreducible representation. Our final expression for $h_{ij}$ reads
\bea
h_{ij}&=&\partial^{k}\epsilon_{ikab}\phi^{ab}_{\ \ j}+\partial^{k}\epsilon_{jkab}\phi^{ab}_{\ \ i}-8kf^{2}Q_{ikj}^{\ \ \ k}\nonumber\\
&&+\partial_{i}u_{j}+\partial_{j}u_{i}-\frac{2}{3}kf^{2}\delta_{ij}P_{mn}^{\ \ mn}\nonumber \\
&=&\partial^{k}\epsilon_{ikab}\phi^{ab}_{\ \ j}+\partial^{k}\epsilon_{jkab}\phi^{ab}_{\ \ i}+\partial_{i}u_{j}+\partial_{j}u_{i}\nonumber\\
&&-8kf^{2}P_{ikj}^{\ \ \ k}+\frac{4}{3}kf^{2}\delta_{ij}P_{mn}^{\ \ mn}.
\eea

Now we shall determine the invariances of the prepotential $\phi_{ijk}$. They are defined by the equation

\bea
\delta h_{ij}&=&\partial_{i}\xi_{j}+\partial_{j}\xi_{i}-2kf^{2}\delta_{ij}\xi\nonumber\\
&=&\partial^{l}\epsilon_{ilab}\delta\phi^{ab}_{\ \ j}+\partial^{l}\epsilon_{jlab}\delta\phi^{ab}_{\ \ i}+\partial_{i}\delta u_{j}+\partial_{j}\delta u_{i}\nonumber\\
&&-8kf^{2}\delta P_{ikj}^{\ \ \ k}+\frac{4}{3}kf^{2}\delta_{ij}\delta P_{mn}^{\ \ mn}. 
\eea
Substituting for $\delta P_{abcd}$ according to (\ref{invPP}) one gets
\bea
\partial^{l}\epsilon_{ilab}\delta\phi^{ab}_{\ \ j}+\partial^{l}\epsilon_{jlab}\delta\phi^{ab}_{\ \ i}&=&\partial_{i}(\xi_{j}-\delta u_{j}+8kf^{2}\chi_{jb}^{\ \ b})\nonumber\\
&+&\partial_{j}(\xi_{i}-\delta u_{i}+8kf^{2}\chi_{ib}^{\ \ b})\nonumber\\
&-&8kf^{2}\partial^{l}\chi_{jli}-8kf^{2}\partial^{l}\chi_{ilj}\nonumber\\
&-&\frac{16}{3}kf^{2}\delta_{ij}\partial^{l}\chi_{lb}^{\ \ \ b}.\nonumber\\ \label{gaugeconlam}
\eea
It is useful to dualize in the antisymmetric pair of indices of $\chi_{abc}$ (and then project onto its (2,1) symmetry)
\[
\chi_{abc}=2\epsilon_{abxy}\tilde{\chi}_{xyc}+\epsilon_{cbxy}\tilde{\chi}_{xya}-\epsilon_{caxy}\tilde{\chi}_{xyb}
\]
so (\ref{gaugeconlam}) takes the form
\bea
&&\partial^{l}\epsilon_{il}^{\ \ ab}(\delta\phi_{abj}+16kf^{2}(\tilde{\chi}_{jba}+\tilde{\chi}_{abj}))\nonumber\\
&+&\partial^{l}\epsilon_{jl}^{\ \ ab}(\delta\phi_{abi}+16kf^{2}(\tilde{\chi}_{iba}+\tilde{\chi}_{abi}))\nonumber\\
&=&\partial_{i}(\xi_{j}-\delta u_{j}+16kf^{2}\epsilon_{jbxy}\tilde{\chi}^{xyb})\nonumber\\
&+&\partial_{j}(\xi_{i}-\delta u_{i}+16kf^{2}\epsilon_{ibxy}\tilde{\chi}^{xyb}).\nonumber\\\label{deduc}
\eea
From the previous expression one finally deduces
\bea
\delta\phi_{abc}&=&\partial_{a}S_{bc}-\partial_{b}S_{ac}+\partial_{a}A_{bc}-\partial_{b}A_{ac}+2\partial_{c}A_{ba}\nonumber\\
&&+B_{\left[a\right.}\delta_{\left.b\right]c}-16kf^{2}(\tilde{\chi}_{cab}+\tilde{\chi}_{abc})\nonumber\\
\eea
and
\bea
\delta u_{i}&=&\xi_{i}+16kf^{2}\epsilon_{ibxy}\tilde{\chi}^{bxy}-2\partial^{l}\epsilon_{il}^{\ \ ab}A_{ba}.
\eea
We observe the appearance of terms in $\tilde{\chi}_{ijk}$, corresponding to an invariance of the prepotential $P_{ijkl}$, as a part of the transformations of the prepotential $\phi_{ijk}$ as well:  the introduction of a cosmological constant mixes the invariance parameters of the prepotentials in this particular manner. The fact that this phenomenon does not occur in four dimensions may be attributed to its especially symmetric character.





\subsection{Action in the prepotential formalism}

Having solved the constraints, we can now write the action principle in terms of the prepotentials by direct substitution in (\ref{actionlambda1}). This yields

\bea
&&S[\phi_{ijk},P_{abcd}]=\nonumber\\
&&\int dt\;d^{4}x\left[2\epsilon^{imab}\epsilon^{jncd}\epsilon_{ilxy}\partial_{m}\partial_{n}P_{abcd}\partial^{l}\dot{\phi}^{xy}_{\ \ j}\right.\nonumber\\
&&+\frac{32}{3}k\dot{P}\partial_{a}\partial_{b}P^{ab}-8k\dot{P}_{ij}\partial_{a}\partial_{b}P^{iajb}\nonumber\\
&&+8kf^{2}\epsilon^{jlab}\partial^{i}\partial_{l}\phi_{ab}^{\ \ k}\partial_{i}P_{jk}-8kf^{2}\epsilon_{jlab}\partial_{i}\partial^{l}\phi^{abi}\partial_{k}P^{kj}\nonumber\\
&&+\frac{72}{9}k^{2}f^{2}\partial_{j}P\partial^{j}P+32k^{2}f^{2}\partial^{i}P_{ik}\partial_{j}P^{jk}\nonumber\\
&&-16f^{2}k^{2}\partial_{i}P_{jk}\partial^{i}P^{jk}-\frac{64}{3}k^{2}f^{2}\partial_{i}P\partial_{j}P^{ij}\nonumber\\
&&-\left(f^{-4}(R_{ij}[P]R ^{ij}[P]-\frac{7}{27}R^{2}[P])\right.\nonumber\\
&&\left.\left.+f^{2}(2E_{ijk}[\phi]E^{ijk}[\phi]-\frac{3}{2}E_{i}[\phi]E^{i}[\phi])\right)\right].\nonumber\\\label{actioncomplete}
\eea
Here $R_{ij}$ and $E_{ijk}$ are defined as in \cite{me5}: they correspond to the  respective contractions of the tensors

\be
R_{ijklmn}=18\partial_{[i}P_{jk][lm,n]}
\ee
and
\be
E_{ijkmn}=6\partial_{[n}\partial_{[i}T_{jk]m]}
\ee
regarded as curvatures for the prepotentials. We see that, in addition to the terms that appear when $\Lambda=0$, there are new terms proportional to $k$ (a power of $\Lambda$) with two derivatives of the prepotentials. Contrarily to the situation in four dimensions, one can not get rid of these terms even after a redefinition by a power of $f$ in the prepotentials (for instance, the terms proportional to $k$ with time derivatives can not be written as a total time derivative). This may again be interpreted as an indication of a special character of the four dimensional case: the complications appearing in higher dimensions happen to vanish, which ultimately gives rise to the $SO(2)$ symmetry acting on the prepotentials. 

\medskip






One may now wonder whether it is possible to derive (\ref{actioncomplete}) from some suitable deformation of the Curtright action once the corresponding constraints are solved. Since one has no prior knowledge of any such deformation, this may in fact be regarded as a definition: the deformed theory, dual to the Einstein-Hilbert action linearized around a de Sitter background (\ref{actionADM2}), must be such that it takes the form (\ref{actioncomplete}) after the resolution of the constraints. However, this definition does not provide any indications about how to construct the deformed action. For such purpose, it seems more convenient to restrict ourselves to a particular gauge in which the terms in (\ref{actioncomplete}) proportional to $k$ are not present: the Hamiltonian would then take the same form as in the case $\Lambda=0$ (up to factors of $f$) and thus one could introduce a dual pair of canonical variables following the steps described in \cite{me5}. We shall see in next section that such a gauge does exist, and the gauge-fixed deformation of the Curtright action will be derived. This suffices in order to show the novel qualitative feature of the dual action principle: its non-local character.



\section{Deformed action}

\setcounter{equation}{0}

The aim of this section is to show how the prepotential formulation previously developed turns out to be useful in the construction of a dual theory satisfying the properties we expect for a deformation of the Curtright action. In order to simplify the analysis, we shall work in the gauge where $\hat{p}=0$ (we see from (\ref{cantra}) that this condition is satisfied, in particular, in the transverse-traceless gauge of \cite{Abbott}). In this gauge, the scalar constraint (\ref{Cscalar}) takes the same form as in the case with no cosmological constant: this leads to the vanishing of the terms proportional to $k$ and $k^{2}$ in (\ref{actioncomplete}), and simplifies the construction of the deformed action. 


\medskip

Thus we shall set

\bea
\hat{p}^{ij}&=&a^{ij}+\delta \hat{p}^{ij}=\hat{a}^{ij}-\partial^{i}\partial^{j}\xi+\delta^{ij}\Delta\xi\nonumber\\
h_{ij}&=&b_{ij}-2kf^{2}\xi\label{gchoice}
\eea
and $a=0$. Clearly this can be achieved through the gauge choice $\xi=\frac{1}{3}\Delta^{-1}\hat{p}$. Now the constraints read as in the case with no cosmological constant \cite{me5}

\bea
\partial_{j}a^{ij}&=&0\\
\Delta b-\partial^{i}\partial^{j}b_{ij}&=&0
\eea
so they may be solved as follows

\bea
a^{ij}&=&f^{-2}\partial^{k}\partial^{l}\epsilon_{ikab}\epsilon_{jlcd} P^{abcd}\label{redefa}\\
b_{ij}&=&f^{2}(\partial^{l}\epsilon_{ilab}\phi^{ab}_{\ \ j}+\partial^{l}\epsilon_{jlab}\phi^{ab}_{\ \ i})+\partial_{i}u_{j}+\partial_{j}u_{i}.\label{redefh}\nonumber\\
\eea
We note that $u_{i}$ is such that $b=2\partial^{m}u_{m}$, and it may actually be dropped for it has the form of a gauge transformation. The prepotentials have been redefined by powers of $f$ for future convenience. 

\medskip

By substituting in the action (\ref{actionlambda1}) one gets 
\bea
&&S[P_{ijkl},\phi_{abc}]=\nonumber\\
&&\int dt\;d^{4}x\; \left[2\partial_{m}\partial_{k}\epsilon^{imnp}\epsilon^{jkst}P_{npst}\partial^{l}\epsilon_{ilab}\dot{\phi}^{ab}_{\ \ j}\right.\nonumber\\
&&-\left(f^{-4}(R_{ij}[P]R^{ij}[P]-\frac{7}{27}R^{2}[P])\right.\nonumber\\
&&+\left.\left.f^{2}(2E^{ijk}[\phi]E_{ijk}[\phi]-\frac{3}{2}E_{i}[\phi]E^{i}[\phi])\right)\right].\label{actionprepC}\nonumber\\
\eea
We observe that, because of our gauge choice (\ref{gchoice}), the terms proportional to $k$ and $k^{2}$ in (\ref{actioncomplete}) are not present any longer, and thus the action has the same form as in the absence of a cosmological constant (up to some powers of $f$ in the Hamiltonian). In this regard, the redefinition of the prepotentials implicit in (\ref{redefa}) and (\ref{redefh}) by powers of $f$ is crucial, for it permits the cancellation of the contributions derived from the term $2kp^{ij}h_{ij}$ in the Hamiltonian (\ref{Hamii}) --owing to a term of opposite sign produced by the time derivative in the kinetic term.

\medskip

In order to construct the dual theory one now introduces the canonical pair of dual variables defined as follows \cite{me5}:

\bea
\hat{t}^{ijk}&=&-\frac{2}{3}f^{-2}\partial_{l}\left[2\epsilon^{klab}P_{ab}^{\ \ ij}+\epsilon^{ilab}P_{ab}^{\ \ kj}-\epsilon^{jlab}P_{ab}^{\ \ ki}\right]\nonumber\\
\hat{\pi}_{ijk}&=&f^{2}\epsilon_{ijmn}\epsilon_{krst}\partial^{m}\partial^{r}\phi^{stn}.\label{definitiondual}
\eea
The action
\bea
S[\hat{t}_{ijk},\hat{\pi}_{ijk},m_{j},m_{ij}]&=&\int d^{5}x\;\left[\hat{\pi}^{ijk}\dot{\hat{t}}_{ijk}-{\cal{H}}\right.\nonumber\\
&&\left.-m_{j}\Gamma^{j}-m_{jk}\Gamma^{jk}\right] \label{actionC3}
\eea
reproduces the form of (\ref{actionprepC}), with the Hamiltonian density
\bea
{\cal{H}}&=&-2k\hat{\pi}^{ijk}\hat{t}_{ijk}+\frac{1}{2}\partial_{i}\hat{t}_{jkl}\partial^{i}\hat{t}^{jkl}+\partial_{i}\hat{t}_{jkl}\partial^{j}\hat{t}^{kil}\nonumber\\
&&-\frac{1}{2}\partial^{k}\hat{t}_{ijk}\partial_{l}\hat{t}^{ijl}+\frac{1}{2}\hat{\pi}_{ijk}\hat{\pi}^{ijk}-\frac{1}{2}\hat{\pi}_{i}^{\ ji}\hat{\pi}^{k}_{\ jk}\nonumber\\
\eea
and the constraints 
\bea
\Gamma^{j}&=&\partial_{i}\partial_{k}\hat{t}^{ijk}\\
\Gamma^{ij}&=&-2\partial_{k}(\hat{\pi}^{ijk}+\hat{\pi}^{kji}).\label{momentumCC}
\eea
The latter need to be imposed so (\ref{definitiondual}) holds, and generate the usual gauge transformations

\bea
\delta t_{ijk}&=&2\partial_{[i}u_{j]k}+2\partial_{[i}v_{j]k}-2\partial_{k}v_{ij},\nonumber\\
\delta\pi^{ijk}&=&\frac{1}{2}(\partial^{j}\partial^{k}\xi^{i}-\partial^{i}\partial^{k}\xi^{j})\nonumber\\
&&+\frac{1}{2}\left(\delta^{jk}(\partial^{i}\partial_{m}\xi^{m}-\Delta\xi^{i})-\delta^{ik}(\partial^{j}\partial_{m}\xi^{m}-\Delta\xi^{j})\right)\nonumber\\
\eea
leaving (\ref{actionC3}) invariant, with $u_{ij}=u_{ji}$, $v_{ij}=-v_{ji}$. 

\medskip

The action (\ref{actionC3}) may be regarded as the dual of the standard action (\ref{actionlambda1}) --written in terms of the `new' variables $(b_{ij},a^{ij})$. In order to obtain an action dual to the original variational principle (\ref{actionlambda2}) --in terms of the `old', graviton canonical variables $(h_{ij},p^{ij})$-- we should find a way to invert the action of the canonical transformation (\ref{cantra}) in the dual picture. This is achieved by writing the generating functional (\ref{genfunc}) in terms of the variables $\hat{t}_{ijk}$ and $\hat{\pi}^{ijk}$ through the inversion formulae \cite{me5}

\be
\phi_{ijk}[\hat{\pi}]=-\frac{1}{2}\Delta^{-1}\hat{\pi}_{ijk}\label{inversion}
\ee
and
\bea
P_{abcd}[\hat{t}]&=&\frac{1}{8}\left[\epsilon_{abij}\partial^{i}\Delta^{-1}\hat{t}_{cd}^{\ \ j}+\epsilon_{cdij}\partial^{i}\Delta^{-1}\hat{t}_{ab}^{\ \ j}\right]-\nonumber\\
&&-\frac{1}{24}\left[\epsilon_{abij}\partial^{i}\Delta^{-1}\hat{t}_{cd}^{\ \ j}+\epsilon_{cdij}\partial^{i}\Delta^{-1}\hat{t}_{ab}^{\ \ j}\right.\nonumber\\
&&+\epsilon_{caij}\partial^{i}\Delta^{-1}\hat{t}_{bd}^{\ \ j}+\epsilon_{adij}\partial^{i}\Delta^{-1}\hat{t}_{bc}^{\ \ j}\nonumber\\
&&\left.+\epsilon_{bcij}\partial^{i}\Delta^{-1}\hat{t}_{ad}^{\ \ j}+\epsilon_{bdij}\partial^{i}\Delta^{-1}\hat{t}_{ca}^{\ \ j}\right].\nonumber\\
\eea
These expressions are suitable for our gauge choice, for they imply $a=b=0$. In addition to the condition $\hat{t}_{ij}^{\ \ j}=0$ (implicit in (\ref{definitiondual})) we can further impose that $\hat{\pi}^{ij}_{\ \ j}=0$: it is consistent with the inversion formula (\ref{inversion}), in the sense that the trace of $\phi_{ijk}$ is pure gauge \cite{me5}, and simplifies the computations.\footnote{We should remind that the gauge parameter that implements a gauge condition on $\pi^{ijk}$ (similarly, on $t_{ijk}$) corresponds to an internal invariance of $h_{ij}$ ($\pi_{ij}$), so the imposition of a gauge condition for the mixed symmetry canonical variables does not modify the gauge parameters of the canonical variables in the standard theory (and vice versa).}

\medskip

Since the canonical transformation (\ref{cantra}) leaves $h_{ij}$ invariant, it is natural to expect in the dual theory the action of the canonical transformation on $\hat{\pi}_{ijk}$ to be the identity map. Therefore we shall set $\hat{\pi}_{ijk}=\pi_{ijk}$. The generating functional takes the form

\bea
F[\pi^{ijk},\hat{t}_{ijk}]&=&\int d^{4}x[-\hat{t}^{ijk}\pi_{ijk}-3k\pi^{ijk}\Delta^{-1}\pi_{ijk}]\label{genFlambda}\nonumber
\eea
where we have dropped a boundary term. We observe that, when expressed in terms of the dual variables, the generating functional depends on the `old' conjugate momentum $\pi^{ijk}$ and the `new' field $\hat{t}_{ijk}$, so the relevant relations are now

\be
t_{ijk}=-\frac{\delta F}{\delta\pi^{ijk}}, \ \ \ \hat{\pi}^{ijk}=-\frac{\delta F}{\delta\hat{t}_{ijk}}.
\ee
The equation for the conjugate momentum yields $\pi^{ijk}=\hat{\pi}^{ijk}$, in agreement with our previous guess. For the Curtright field, one finds the spatially non-local expression
\be
t_{ijk}=-\frac{\delta F}{\delta\pi^{ijk}}=\hat{t}_{ijk}+6k\Delta^{-1}\pi_{ijk}
\ee
The action (\ref{actionC3}) is now expressed in terms of the pair $(t_{ijk}, \pi^{ijk})$:

\bea
S[t_{ijk},\pi^{ijk},m_{i},m_{ij}]&=&\int dtd^{4}x\;\left[\pi^{ijk}\dot{t}_{ijk}-{\cal{H}}\right.\nonumber\\
&&\left.-m_{i}\Gamma^{i}-m_{ij}\Gamma^{ij}\right]\label{SSS}
\eea
We have dropped a total time derivative originating from the kinetic term and recast the Hamiltonian density as the sum

\be
{\cal{H}}={\cal{H}}_{0}+{\cal{H}}_{\Lambda}\label{defH}
\ee
with
\bea
{\cal{H}}_{0}&=&\frac{1}{2}\partial_{i}t_{jkl}\partial^{i}t^{jkl}+\partial_{i}t_{jkl}\partial^{j}t^{kjl}-\frac{1}{2}\partial^{k}t_{ijk}\partial_{l}t^{ijl}\nonumber\\
&&+\frac{1}{2}\pi_{ijk}\pi^{ijk}-\frac{1}{2}\pi_{i}^{\ ji}\pi^{k}_{\ jk}
\eea
the contribution from the free theory and
\bea
{\cal{H}}_{\Lambda}&=&4k\pi^{ijk}t_{ijk}-6k^{2}\pi^{ijk}\Delta^{-1}\pi_{ijk}
\eea
the term carrying the deformation. As in (\ref{actionlambda2}), one distinguishes two contributions to the deformation: one linear and other quadratic in $k$, with the novelty here of the non-local character of the latter. The constraints read
\bea
\Gamma^{j}&=&\partial_{i}\partial_{k}(t^{ijk}-6k\Delta^{-1}\pi_{ijk}-\delta_{ijk})\\
\Gamma^{ij}&=&-2\partial_{k}(\pi^{ijk}+\pi^{kji}).
\eea
The action principle (\ref{SSS}) may then be regarded as a gauge-fixed deformation of the Curtright action in its Hamiltonian form \cite{me5}. Its most prominent feature consists in a spatially non-local character, reflected in the term proportional to $k^{2}$ of the deformed Hamiltonian, in consonance with the main result in \cite{HenBou}. The undeformed theory is recovered in the limit $k\rightarrow 0$. To the best of our knowledge, this is the first instance of a deformation of the Curtright action in the literature.





\section{Conclusions and comments}

We have constructed an action principle, dual to the linearized version of five-dimensional gravity around a de Sitter background, in terms of the mixed symmetry canonical variables associated to the dual graviton. This construction can be naturally regarded as a deformation of the Curtright action corresponding to the introduction of a positive cosmological constant in the Pauli-Fierz theory and relies on the resolution of the corresponding constraints in terms of two prepotentials, which turn out to possess the same mixed symmetries as in the case with no cosmological constant. Though the resolution of these constraints can be carried out with no need to specify any gauge condition, the construction of the dual action is particularly simplified upon a specific gauge choice, where the constraints and the Hamiltonian --after performing a suitable canonical transformation defining new variables $(\hat{h}_{ij}, \hat{\pi}^{ij})$-- take essentially the same form as they do in a Minkowski background (up to time-dependent factors). Consequently, the prepotential formulation of the action principle also takes the same form as in the case $\Lambda=0$ (again, up to time-dependent factors), and this allows us to introduce a canonical pair of dual variables by pertinent inversion formulae, imposing the usual Curtright constraints on them. These, however, are the dual version of the `new', canonically transformed variables $(\hat{h}_{ij}, \hat{\pi}^{ij})$, not of the `old', original ones $(h_{ij}, \pi^{ij})$. After reversing the canonical transformation in the dual picture, one obtains a spatially non-local action principle. This is the main qualitative feature of our construction, sufficient to circumvent no-go results based on the hypotheses of locality and manifest Lorentz invariance.


\medskip




Since we have specified a particular gauge in the derivation of our result, one may wish to deal with the general situation and derive an ungauged deformed action. In this ungauged version of the dual theory, one expects both the Hamiltonian and the constraints to include new terms, in such a way that the ungauged prepotential action (\ref{actioncomplete}) is recovered when the constraints are solved. As we have observed, gauge transformations of the standard canonical variables $(h_{ij}, \pi^{ij})$ do not affect the dual, mixed symmetry canonical variables (and viceversa), so there is no way to determine these additional terms in the dual theory from the knowledge of the gauge parameters in the standard picture. Nevertheless, one can guess some of their properties by simple arguments. For instance, one would expect a deformation in the constraint for $\hat{\pi}^{ijk}$ by terms depending solely on $\hat{t}_{ijk}$, and no deformation in the constraint for $\hat{t}_{ijk}$, thus inverting the situation encountered in the standard picture (a deformation in the constraint for $h_{ij}$ depending only on $\hat{p}^{ij}$ and no deformation in the constraint for $\hat{p}^{ij}$). Moreover, one should be able to recast the deformation of the constraint for $\hat{\pi}^{ijk}$ in the form of a gauge transformation $\delta \hat{t}_{ijk}$ (so it can actually be gauged away), with the parameter depending on $\hat{t}_{ijk}$, and its vanishing should imply $\hat{p}=0$. A strategy to determine the exact form of these additional terms in the ungauged dual action is to consider all possible contributions that satisfy the aforementioned requirements and then substitute in the action in order to adjust their relative coefficients, in such a way that (\ref{actioncomplete}) is recovered in the end. It would be of interest to know the exact form of these additional terms, and see whether they bring about any qualitatively new features.

\medskip


One should also emphasize that the deformed, non-local action we have derived corresponds merely to the introduction of a cosmological constant in the Pauli-Fierz picture, and not to the inclusion of higher order terms in the perturbative expansion of the Einstein-Hilbert action. On the other hand, one may well wonder about the existence of deformations of the Curtright theory corresponding to linearization of the Einstein-Hilbert action around other space-time backgrounds, such as anti de Sitter or, more generally, conformally flat metrics. If such a deformation should be obtained by the method we have described, this would require not only the feasibility of a prepotential resolution of the corresponding linearized constraints (as yet, a subject to be investigated also in four dimensions), but also our ability to recast them, by some change of variables, in the same form as in the flat case: in this way one could establish the Curtright constraints on the dual side and simply introduce the dual pair of canonical variables by the already known inversion formulae, reversing the change of variables on the dual side afterwards. 
\medskip





Apart from these technical issues, our result also rises more conceptual questions, such as the interpretation of a cosmological constant in the dual theory, or the properties that space and time should possess in such an alternative formulation --where the concept of constant curvature in the geometric theory gets translated into spatial non-locality\footnote{A possibility to be considered is whether one can reestablish either locality or manifest space-time covariance by the introduction of auxiliary fields.}. The answer to these questions is certainly far from obvious, and it would constitute a part of the challenging enterprise of interpreting higher dimensional, (pseudo) Riemannian geometry in a dual, non-geometric formulation where the basic objects possess a mixed symmetry character. Since the necessity of such a complicated description of gravity seems hard to justify a priori --especially when we dispose of the elegant, generally covariant description that General Relativity provides-- we can only hope that a further understanding of gravitational duality may be relevant in order to gain definitive insight into the, yet conjectural and rather mysterious, extended Kac-Moody algebra structure.

\medskip

\medskip

\medskip

\medskip

\section*{Acknowledgments} 
I thank M. Henneaux for suggesting the problem addressed in this article and a detailed reading of the manuscript. This work has been supported by the Fondecyt grant N\textordmasculine \; 3160781. The Centro de Estudios Cient\'ificos (CECs) is funded by the Chilean Government through the Centers of Excellence Base Financing Program of Conicyt.

\end{document}